\documentclass{aa}
\usepackage{graphicx}
\begin{document}

\title{The magnetic fields of large Virgo Cluster 
spirals$^*$}

\author {
M. We\.zgowiec\inst{1}
\and M. Urbanik\inst{1}
 \and B. Vollmer\inst{2}
 \and R. Beck\inst{3}
 \and K. T. Chy\.zy\inst{1}
 \and M. Soida\inst{1}
 \and Ch. Balkowski\inst{4}}
\institute{Obserwatorium Astronomiczne Uniwersytetu
Jagiello\'nskiego, ul. Orla 171, 30-244 Krak\'ow, Poland
\and CDS, Observatoire Astronomique de Strasbourg, UMR                    
7550, 11 rue de l'universite, 67000 Strasbourg, France                          
\and Max-Planck-Institut f\"ur Radioastronomie, Auf dem
H\"ugel 69, 53121 Bonn, Germany
\and Observatoire de Paris, GEPI, CNRS, and Universite Paris 7,
5 Place Jules Janssen, 92195 Meudon Cedex, France}

\offprints{M. We\.zgowiec}
\mail{markmet@oa.uj.edu.pl\\
$^*$Based on the observations with the 100-m telescope at Effelsberg
operated by the Max-Planck-Institut f\"ur Radioastronomie (MPIfR) on behalf
of the  Max-Planck-Gesellschaft.}
\date{Received date/ Accepted date}

\titlerunning{The magnetic fields of large Virgo Cluster 
spirals}
\authorrunning{M. We\.zgowiec et al.}

\abstract
{Because of its proximity the Virgo Cluster is an excellent target
for studying interactions of galaxies with the cluster 
environment. Both the high-velocity tidal interactions and 
effects of ram pressure stripping by the intracluster gas can be 
investigated.}
{Optical and/or \ion{H}{i} observations do not always show 
effects of weak interactions between galaxies and their 
encounters with the cluster medium. For this reason we  searched
for possible anomalies in the magnetic  field structure in Virgo
Cluster spirals which could be  attributed to perturbations in
their gas distribution and  kinematics.}
{Five angularly large Virgo Cluster spiral galaxies (NGC\,4501,
NGC\,4438, NGC\,4535, NGC\,4548 and NGC\,4654) were the targets for
a sensitive total power and polarization study using the 100-m
radio telescope in Effelsberg at 
4.85~GHz. For two objects polarization data at 
higher frequencies have been obtained allowing Faraday rotation
analysis.}
{Distorted magnetic field structures were identified in all
galaxies. Interaction-induced magnetized outflows were found in
NGC\,4438  (due to nuclear activity) and NGC\,4654 (a combination
of tidal  tails and ram pressure effects). Almost all objects
(except the  anaemic NGC~4548) exhibit distortions in polarized
radio continuum  attributable to influence of the ambient
gas. For some galaxies they agree with observations of other
species, but sometimes (NGC\,4535) the magnetic field is the only
tracer of the interaction with the cluster environment.} 
{The cluster environment clearly affects the evolution of the
galaxies due to ram pressure and tidal effects. Magnetic 
fields provide a very long-lasting memory of past interactions. 
Therefore, they are a good tracer of weak interactions which 
are difficult to  detect  by other observations. 
Information about motions of galaxies
in the sky plane and their three-dimensional distribution 
can also be obtained.}

\maketitle

\section{Introduction}

Cluster galaxies are known to interact intensively with the 
intracluster medium (ICM). One of the effects is ram pressure 
stripping first discussed by Gunn \& Gott (\cite{gunn}). This 
process is particularly active in central parts of a cluster with
high-density ICM giving rise to ram pressure effects and
stripping of the  outer parts of the gaseous disks of galaxies
(Vollmer et  al.~\cite{vollmer2}, ~\cite{vollmer04a}). This leads
to significant \ion{H}{i}  deficiencies in such objects (Cayatte
et al.~\cite{cayatte},  ~\cite{cayatte2}). These interactions
seem to influence only the neutral and the ionized gas (Chemin et
al.~\cite{chemin}),  while the observations of stellar
populations (Gavazzi et al.~\cite{gavazzi}, ~\cite{gavazzi2}) and
of CO  content (Boselli et al. \cite{boselli}) show no traces of ram
pressure  effects. However, tidal interactions (see Toomre \& 
Toomre~\cite{toomre}) may be also important for cluster galaxies 
as their relative distances are about 10~times smaller than
between field objects. The cluster galaxies can interact
either with the cluster potential (Byrd \& Valtonen~\cite{byrd},
Valluri~\cite{valluri}) or with other cluster galaxies (''galaxy 
harassment'', Moore et al.~\cite{moore}). Special attention should 
be given to the high-velocity tidal encounters which occur almost 
exclusively in the cluster environment. Due to the high relative velocities
between the cluster galaxies a  small impact  parameter is
required  to perturb them efficiently. The probability of such
encounters is rather low. We also note that in contrast to ram
pressure effects, tidal interactions affect both the gas and
stellar content of a galaxy.

The interactions between the interstellar medium of the galaxies 
(ISM) and the ICM are already known to cause strong gas 
compression effects, observed as \ion{H}{i} ridges in the outskirts of 
galactic disks (Cayatte et al.~\cite{cayatte}). Observations of the 
polarized emission in perturbed galaxies are known to provide a 
sensitive tool to trace the anomalous gas motions and  
compression effects even when the disturbances are hardly 
visible in other species (Soida et al.~\cite{soida2}). Moreover, 
for cluster galaxies a unique possibility to determine the 
directions of the gas flows in the sky plane arises, adding a new 
dimension to the radial velocity studies 
(Urbanik~\cite{urbanik2}).

 Due to its proximity and a high content of spiral galaxies, the 
Virgo Cluster is the best target to observe the effects of both kinds 
of interactions. It is the nearest cluster of galaxies, situated at the 
distance of only 17~Mpc, which enables a sufficient resolution 
even with single-dish radio telescopes. The cluster covers an area of approx.
10$\times$10$\degr$ (3$\times$3~Mpc in the sky plane) 
with even larger extent towards the southern extension. The main subcluster 
concentrates around M\,87 and is  characterized by a high density of 
the ICM as seen in X-ray maps of B\"ohringer (\cite{bohringer}). 
The southern extension of the cluster forms another subcluster 
concentrated around the giant elliptical galaxy M\,49. It is smaller, 
less dense and associated with a distinct concentration of hot gas 
seen in the X-ray maps (B\"ohringer~\cite{bohringer}). More 
distinct subclusters have been identified by Gavazzi (\cite{gav77}).

So far, detailed information on possible effects of the cluster   
environment on the galactic magnetic fields is restricted to
three Virgo  Cluster objects. In the first one -- NGC\,4254 -- only
weakly affected by  the ambient gas, a strongly polarized region
on the disk side facing the cluster core has
been found (Chy\.zy et al.~\cite{chyzy}, Soida  et
al.~\cite{soida}). In the highly inclined NGC\,4522 the polarized
emission is asymmetric, with extension coinciding with
extraplanar H$\alpha$ emission (Vollmer at al.
\cite{vollmer04b}). The third galaxy -- NGC\,4569 -- was found to 
possess large vertical polarized radio lobes, at least one of
which also shows signs of external compression (Chy\.zy et 
al.~\cite{chyzy2}).  

In order to obtain comprehensive information on global magnetic
field structures of cluster spirals undergoing various types of
interactions with the cluster environment, we performed the first
systematic study of radio polarization of   five angularly large
Virgo Cluster spirals with the Effelsberg 100-m radio telescope.  
To search for large scale outflows or radio tails we observed all
objects in total power   and polarization at 4.85~GHz with a beam
of $2\farcm 5$, (1\arcmin\, corresponds to about 5~kpc at
the distance to the Virgo Cluster).  Objects with radio disks
bright enough were additionally   observed at 10.45~GHz or at 
8.35~GHz (beams of $1\farcm 13$, and $1\farcm 5$, resp.),
in order to obtain more details of the magnetic field structure and Faraday
rotation properties. We chose the following galaxies:

\begin{itemize}

\item NGC\,4438  -- a strongly stripped galaxy in the Virgo
Cluster centre (Chemin et al.~\cite{chemin4438}), also perturbed  by tidal forces from its
companion (Vollmer et al.~\cite{vollmer4}).

\item NGC\,4501 -- a galaxy showing
significant (but not extreme)  effects of gas stripping
(Cayatte et al.~\cite{cayatte}, ~\cite{cayatte2}; Chemin et al.~\cite{chemin}) and a strong
asymmetry in the \ion{H}{i} distribution. It is situated close to an
X-ray cloud  found by B\"ohringer et al. (\cite{bohringer}). 

\item NGC\,4535 -- a relatively unperturbed, optically symmetric
grand-design spiral  in the southern extension of the Virgo
Cluster with a fairly symmetric \ion{H}{i} distribution (Cayatte et
al.~\cite{cayatte}, Chemin et al.~\cite{chemin}). 

\item NGC\,4548 -- an anaemic galaxy with a strong \ion{H}{i} deficiency 
and a low star formation level (Cayatte et
al.~\cite{cayatte},~\cite{cayatte2}; Chemin et al.~\cite{chemin}).

\item NGC\,4654 -- a mildly stripped galaxy (Cayatte et
al.~\cite{cayatte}, ~\cite{cayatte2}; Chemin et al.~\cite{chemin}) with cometary appearance 
and an \ion{H}{i} tail probably caused by joint effects of tidal interactions with 
NGC\,4639 and ram pressure stripping (Vollmer~\cite{vollmer5}, 
Phookun \& Mundy~\cite{phookun}). 

\end{itemize}

\section{Observations and data reduction}

The observations were performed between March 2001 and May 2003 using the 100-m Effelsberg  radio
telescope of the Max-Planck-Institut f\"ur
Radioastronomie\footnote{http://www.mpifr-bonn.mpg.de}  (MPIfR)
in Bonn. The basic astronomical 
properties of the observed objects and of our observations are summarized in
Table~\ref{objects}.

\begin{table*}[t]
\caption{Basic astronomical properties and parameters of radio observations 
of studied galaxies} 
\begin{center}
\begin{tabular}{cccccclc}
\hline
\vspace{1pt} NGC &\vspace{1pt} Morph. &\vspace{-9pt} Optical position$^{\rm 
a}$ 
&\vspace{1pt} Incl.$^{\rm a}$ &\vspace{1pt} Pos. & Dist. to 
&\vspace{1pt} Number 
&\vspace{1pt} r.m.s. in final map \\
& type$^{\rm a}$& \hspace{5pt} $\textstyle\alpha_{2000}$ \hspace{30pt} 
$\textstyle\delta_{2000}$ & [\degr] & ang.$^{\rm a}$[\degr] & Vir A [\degr]&of cov. & [mJy/b.a.] \\
& & & & & & & TP \hspace{15pt} PI\\
\hline
4438 & Sa & 12$^{\rm h}$27$^{\rm m}$45\fs9 \hspace{1pt} +13\degr 00\arcmin 
32\farcs 2 
& 90 & 27 & 0.9 & 25 & \hspace{2pt} 0.7 \hspace{15pt} 0.07\\
4501 & Sb & 12$^{\rm h}$31$^{\rm m}$59\fs3 \hspace{1pt} +14\degr 25\arcmin 
13\farcs 7 
& 60 & 138 & 2 & 12  & \hspace{2pt} 0.9 \hspace{15pt} 0.09\\
& & & & & & 23$^{\rm b}$ & \hspace{6pt} 0.4$^{\rm b}$ \hspace{11pt} 0.17$^{\rm b}$\\
4535 & SBc & 12$^{\rm h}$34$^{\rm m}$20\fs4 \hspace{1pt} +08\degr 
11\arcmin 52\farcs 3 
& 41.3 & 180 & 4.3 & 10 & 0.7 \hspace{16pt} 0.1\\
4548 & SBb & 12$^{\rm h}$35$^{\rm m}$26\fs4 \hspace{1pt} +14\degr 
29\arcmin 47\farcs 0 
& 35 & 150 & 2.4 & 11 & \hspace{2pt} 0.8 \hspace{15pt} 0.09\\
4654 & SBc & 12$^{\rm h}$43$^{\rm m}$57\fs0 \hspace{1pt} +13\degr 
07\arcmin 34\farcs 3 
& 57.7 & 122 & 3.3 & 15 &\hspace{2pt} 0.6 \hspace{15pt} 0.06\\
& & & & & & 33$^{\rm c}$ & \hspace{6pt} 0.25$^{\rm c}$ \hspace{7pt} 0.03$^{\rm c}$\\
& & & & & & 31$^{\rm b}$ & \hspace{6pt} 0.6$^{\rm b}$ \hspace{11pt} 0.14$^{\rm b}$\\
 \hline
\end{tabular}
\label{objects}
\end{center}
$^{\rm a}$ taken from HYPERLEDA database -- http://leda.univ-lyon1.fr -- see
Paturel et al.~\cite{leda}\\
$^{\rm b}$ at 10.45 GHz\\
$^{\rm c}$ at 8.35 GHz
\end{table*}

All galaxies were observed at 4.85~GHz  using the two-horn system
(with horn separation of 8\arcmin) in the secondary 
focus of the radio telescope (see Gioia et
al.~\cite{gioia}). NGC\,4501 and NGC\,4654 have been additionally
observed at 10.45~GHz with a  four-horn (3\arcmin between adjacent horns) 
system (described by Schmidt et al.~\cite{schmidt}). 
NGC\,4654 has additionally been
measured at 8.35~GHz with a single horn. Each horn was equipped
with two total power  receivers and an IF polarimeter resulting
in 4 channels containing the Stokes parameters I (2 channels), Q and U. The
telescope pointing was  corrected by performing cross-scans of a
bright point source close to  the observed galaxy. The flux
density scale was established by  mapping point sources 3C\,138
and 3C\,286.  

The data reduction  was performed using the NOD2 data reduction 
package (Haslam~\cite{haslam}). At 4.85~GHz and at 10.45~GHz 
(dual or multibeam systems used) a  number of coverages in the 
azimuth-elevation frame were obtained for each galaxy, as 
indicated in Table~\ref{objects}.  By combining the information 
from appropriate horns, using the "software beam-switching" 
technique (Morsi \& Reich~\cite{morsi})  followed by a
restoration  of total intensities we obtained  for each coverage
the I, Q, U maps of the galaxy. The restoration method and
its limitations are described in Emerson et
al.~(\cite{emerson2}). At 8.35~GHz  (single horn) we performed 
scans alternatively along the R.A. and  Dec. directions, hence no
restoration was necessary.

All coverages were combined using the spatial frequency  
weighting method (Emerson \& Gr\"ave~\cite{emerson}) yielding 
the final maps of total power, polarized intensity, polarization 
degree and polarization position angles. To show the structure of 
the magnetic field projected on the sky plane we use apparent 
B-vectors defined as E-vectors rotated by $90\degr$. In order to 
remove spatial frequencies corresponding to noisy structures 
smaller than the beam size, a digital filtering process was applied 
to the final maps.

In objects showing weak polarized features in the vicinity 
of a strong unpolarized nuclear source (e.g. NGC\,4438, 
Sect.~\ref{results}) we 
used the  diagrams of instrumentally polarized signal kindly
supplied by Dr H. Rottmann (see also Rottmann et
al.~\cite{rottmann}) and converted the instrumental U and Q
patterns to the form of maps of polarized intensity, scaled to
the same amplitude of total power as the unpolarized source and 
compared them to our observations, in order to estimate the level of 
influence of the instrumental polarization on our data (see Sect.~\ref{4438}). 

\section{Results}
\label{results}

Sections 3.1-3.6 present maps of individual galaxies. Their
integrated data are summarized in Table~\ref{obdata} at the end
of Sect.~3. The data includes total power and polarized flux
densities aquired by integration in 
polygonal areas encompassing all of the radio emission.  
From these data polarization degrees have been 
obtained, also presented in Table~\ref{obdata}. 

\subsection{NGC\,4438}
\label{4438}

NGC\,4438 is located close to the cluster centre, about
$0\fdg 9$ (270\,kpc in the sky plane) from M\,87 (Virgo A). It
interacts tidally with its close companion NGC\,4435 (Combes et
al.~\cite{combes}). 

\begin{figure}[ht]
 \begin{center}
  \resizebox{8cm}{!}{\includegraphics{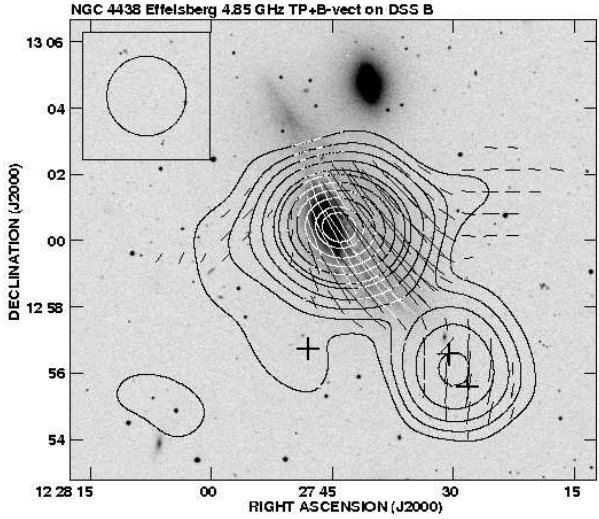}}
 \end{center}
 \caption{ 
The total power map of NGC\,4438 at 4.85~GHz with
apparent B-vectors of polarized intensity overlaid onto the DSS
blue image.  The contours are 3, 6, 10, 16, 25, 30, 40, 50, 60,
70, 75 $\times$ 0.71~mJy/b.a.  and a vector of $1\arcmin$ 
length corresponds to the polarized intensity of 0.7~mJy/b.a. The map
resolution is $2\farcm 5$. The companion galaxy NGC\,4435 is
visible in the north. The beam size is shown in the top left corner of the figure.
Crosses indicate the positions of the background sources (see text).}
\label{4438tp}
\end{figure}

\begin{figure}[ht]
 \begin{center}
  \resizebox{8cm}{!}{\includegraphics{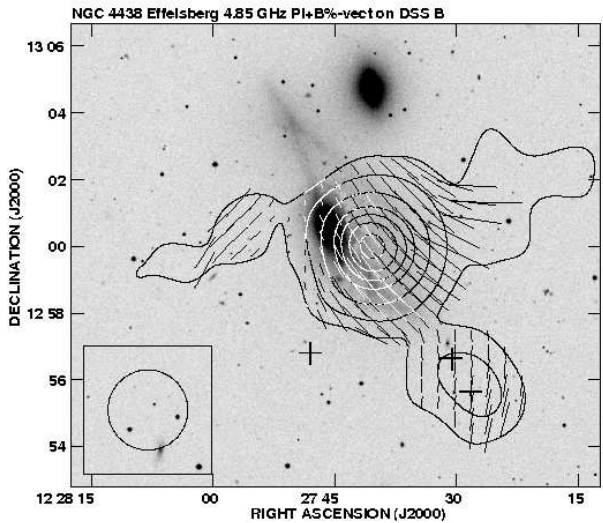}}
 \end{center}
 \caption{ 
The map of polarized intensity of NGC\,4438  at  4.85~GHz with 
apparent B-vectors of polarization degree overlaid onto  the DSS 
blue image. The contours are 3, 8, 16, 25, 30, 35, 40 $\times$ 
0.07~mJy/b.a. and a vector of $1\arcmin$ length corresponds to 
the polarization degree of 7.5\%. The map resolution is $2\farcm 5$. 
The beam size is shown in the bottom left corner of the figure. 
Crosses indicate the positions of the background sources (see text).}
\label{4438pi}  
\end{figure}

The peak of total power coincides with the position of the
galaxy's  centre (Fig.~\ref{4438tp}). The polarization  B-vectors
are oriented  here parallel to the disk. The peak of polarized
intensity is, however, located outside of the bright star-forming
disk  (Fig.~\ref{4438pi}). It is displaced from the optical
centre towards the southwest by about 55~arcseconds (4.6~kpc). 
At this position the optical images  show a complex of dust lanes
that indicate strong compression  effects. Other ISM tracers
(\ion{H}{i}, H$\alpha$, CO, FIR, X-rays) also show emission to
the west of NGC\,4438 (e.g. Kenney et al.~\cite{kenney}, Machacek
et al.~\cite{machacek}, Chemin et al.~\cite{chemin4438}).  We
find that most of polarized  emission ($\simeq$ 74\%) comes from
the western side of the major axis.

The tail extending towards the southwest visible on both total
power and polarized intensity maps (Figs.~\ref{4438tp}
\&~\ref{4438pi}) is due to a blend of two unresolved background
sources, well visible in the map by Condon (\cite{condon}) at
1.49~GHz. These are the NVSS radio sources J$122730.6+125629$ and
J$122728.5+125535$ with total flux densities (at 1.4~GHz) of
38.52~mJy and 19.76~mJy,  respectively (Condon~\cite{nvss}).
Another weak radio source causes the total power extension to the
south.  It is J$122747.6+125647$ with total flux  density of
2.48~mJy at 1.4~GHz. 

 Taking into account possible confusion effects we state that 
NGC\,4438 shows evidence for low surface brightness extraplanar 
features on either side of the galaxy disk:  R.A.$_{2000}=\rm
12^h27^m48\fs5$, Dec.$_{2000}=13\degr00\arcmin59\arcsec$,   and
R.A.$_{2000}=\rm 12^h27^m26^s$,  Dec.$_{2000}=\rm
13\degr01\arcmin30\arcsec$. In total power the effect is less 
pronounced than in polarization because of a considerably higher
noise level (Tab.~\ref{objects}). It is however not due to
sidelobe effects. We used the response of the Effelsberg radio
telescope to a point source of the same intensity as in NGC\,4438
to simulate the map made from a number of coverages with the
range of parallactic angles similar to that for this galaxy. 
Around the main beam response we obtained a weak sidelobe ring
with the brightness of 0.5 -- 0.7~mJy/b.a.. This is three  times
below the level of our first contour in Fig.~\ref{4438tp} thus, 
the telescope sidelobes cannot be entirely responsible for the
total power extensions in NGC\,4438.

The extraplanar extensions are much better visible in
polarization (Fig.~\ref{4438pi}) and extend to at least 25\,kpc
(5\arcmin) distance from the galaxy's optical centre. No emission 
is visible in these regions in the Condon's and NVSS maps. The
polarization B-vectors are highly  inclined to the disk. To check
for  possible contribution of spurious instrumental  polarization
from a strong, unpolarized central source to these  structures we
used the aforementioned instrumental polarization patterns.  We
compared the cross-sections of observed and instrumentally 
polarized signal at various position angles. We found that the
peaks  of instrumental polarization with an amplitude of about 
0.3~mJy/b.a. exist at all position angles at the distance of some
70\arcsec \, (roughly half of the beam), dropping  to almost zero
just beyond 3\arcmin \, from the centre. No instrumentally
polarized signal stronger than 0.07~mJy/b.a. (1$\sigma$ of the
polarized intensity map) has been found at distances greater than
one beam size from the centre. In contrast to that, in NGC\,4438
the polarized signal stronger than 3$\sigma$ continues up to 2
beam sizes (5\arcmin) away from the centre for position angles
between 90\degr \, and 120\degr. This agrees well with the
orientation of extraplanar structures found  by Hummel
\&  Saikia~(\cite{hummel}). The apparent B-vectors are  aligned
with the discussed extensions as also found in NGC\,4569, another 
Virgo Cluster galaxy with large-scale outflows (Chy\.zy et 
al.~\cite{chyzy2}). By all these reasons we believe that the 
vertical extensions in NGC\,4438 cannot be due to instrumental 
polarization and are the real structures. 

No radio emission stronger than 2~mJy/b.a. in total power 
and 0.2~mJy/b.a. in polarized intensity has been detected at the
position of an S0 companion of NGC\,4438 -- NGC\,4435. This is
not surprising: though this galaxy contains some CO (Vollmer et
al.~\cite{vollmer4}), the mass of molecular gas is 20 times
smaller than that of NGC\,4438. NGC\,4435 also shows a very weak 
H$\alpha$ emission extending only few arcsec from the centre 
(Kenney et al.~\cite{kenney}). This indicates a low content of
young stars (see also Boselli et al.~\cite{bos4435}) and of X-ray
emitting gas (Burstein et al.~\cite{x4435}) produced by
supernovae responsible for accelerating the radio-emitting 
cosmic-ray electrons.

\subsection{NGC\,4501}

NGC\,4501 is located at the distance of 2$\degr$ (0.6\,Mpc in the sky plane) from the cluster 
core.  As it is poorly resolved at 4.85~GHz we discuss its radio 
structure  using our observations at 10.45~GHz with almost
twice  better resolution.

\begin{figure}[ht]
\begin{center}
\resizebox{8cm}{!}{\includegraphics{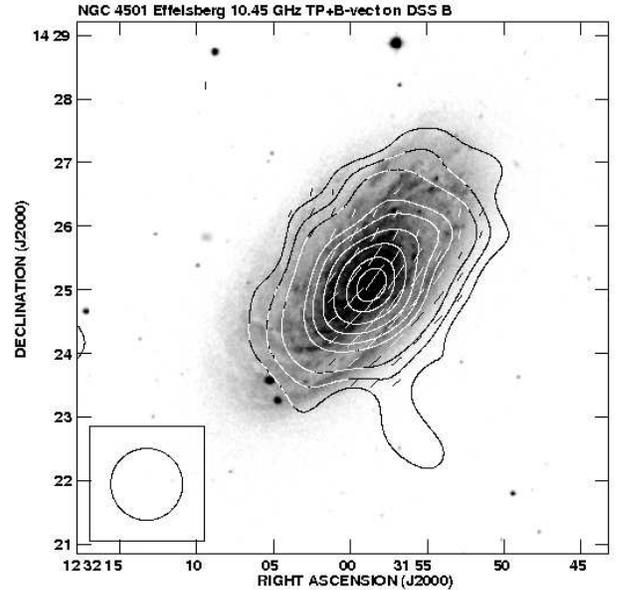}}
\end{center}
\caption{
The total power map of NGC\,4501 at 10.45 GHz with
apparent B-vectors of polarized intensity overlaid onto the
DSS blue image. The contours are 3, 5, 8, 16, 20, 25, 30, 35,
38 $\times$ 0.44~mJy/b.a. and a vector of $1\arcmin$ length
corresponds to the polarized intensity of 5~mJy/b.a. The map
resolution is $1\farcm 13$. The beam size is shown in the bottom left corner
of the figure.
}
\label{4501tp}
\end{figure}

The total power emission at 10.45 GHz from this galaxy 
(Fig.~\ref{4501tp}) is symmetric and  coincides with the optical 
disk. The only extension outside the disk at R.A.$_{2000}=\rm
12^h31^m56^s$,  Dec.$_{2000}=14\degr22\arcmin30\arcsec$  is
due to a weak background source visible in the NVSS map. 

\begin{figure}[ht]
\begin{center}
 \resizebox{8cm}{!}{\includegraphics{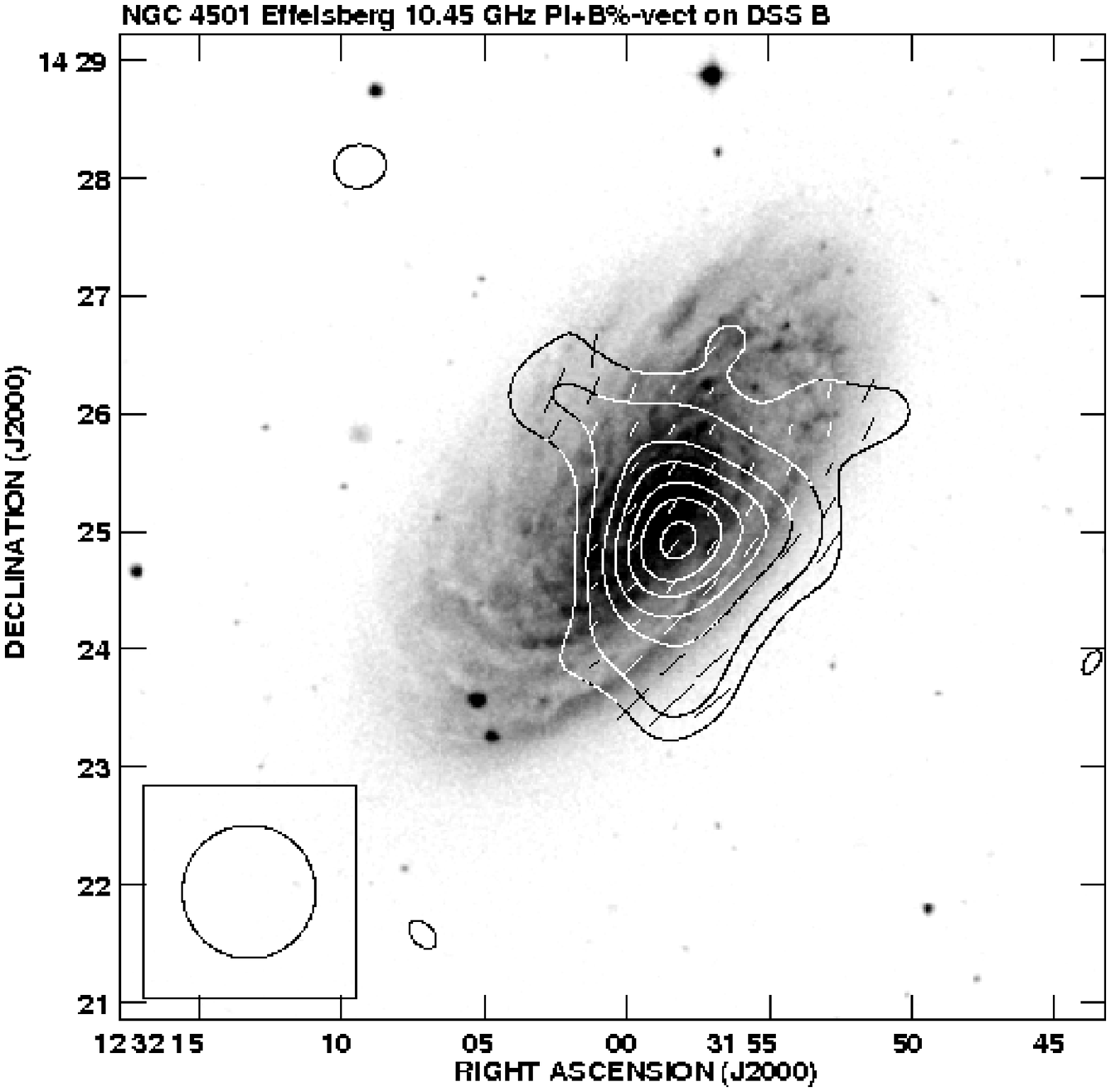}}
 \end{center}
 \caption{
The map of polarized intensity of NGC\,4501 at 10.45  GHz with 
apparent B-vectors of polarization degree overlaid onto  the DSS 
blue image. The contours are 3, 5, 8, 10, 12, 13.5, 15 $\times$ 
0.17~mJy/b.a. and a vector of $30\arcsec$ length corresponds to 
the polarization degree of 50\%. The map resolution  is $1\farcm 
13$.  The beam size is shown in the bottom left corner of the figure. 
}
\label{4501pi}
\end{figure}

The polarized emission is strongly shifted towards the southwest 
(Fig.~\ref{4501pi}), over  85\% of the polarized flux
originates on the SW side of the major axis. The degree of
polarization is quite high in this region, reaching some  20\%.

\subsection{NGC\,4535}
\label{4535}

NGC\,4535 is a grand-design spiral galaxy located in the southern extension of
the Virgo Cluster at the distance of $4\fdg 3$ (1.29\,Mpc in the sky plane) from Virgo\,A. 

\begin{figure}[ht]
 \begin{center}
  \resizebox{8cm}{!}{\includegraphics{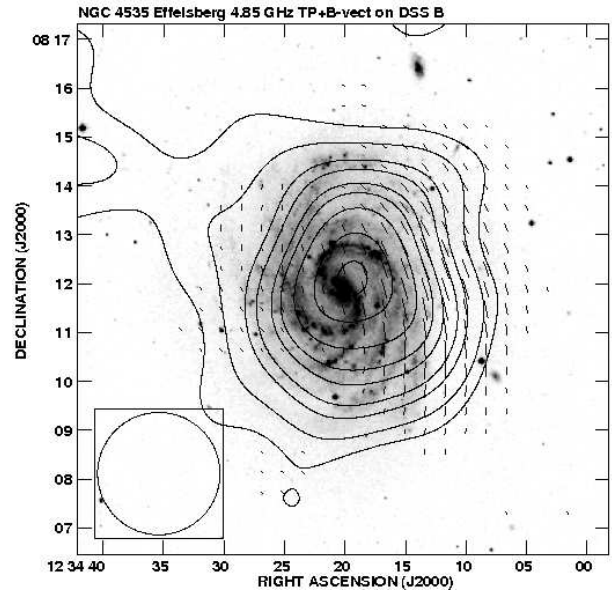}}
 \end{center}
 \caption{ 
The total power map of NGC\,4535 at 4.85 GHz with apparent B-vectors of 
polarized intensity overlaid onto the DSS  blue image. The 
contours are 3, 5, 8, 10, 13, 15, 18, 22, 25 $\times$ 0.68~mJy/b.a. and 
a vector of $1\arcmin$ length corresponds  to the polarized 
intensity of 3.75~mJy/b.a. The map resolution is  $2\farcm 5$. The 
beam size is shown in the bottom left corner of the figure. 
}
\label{4535tp}
  \end{figure}

Optical images show that NGC\,4535 has a very regular 
spiral structure and shows a quite symmetric \ion{H}{i} distribution (Cayatte 
et al.~\cite{cayatte}). The emission in total intensity is distributed in a 
fairly symmetric manner as well. Its peak is roughly situated at the 
galaxy's centre (Fig.~\ref{4535tp}).       

\begin{figure}[ht]
 \begin{center}
  \resizebox{8cm}{!}{\includegraphics{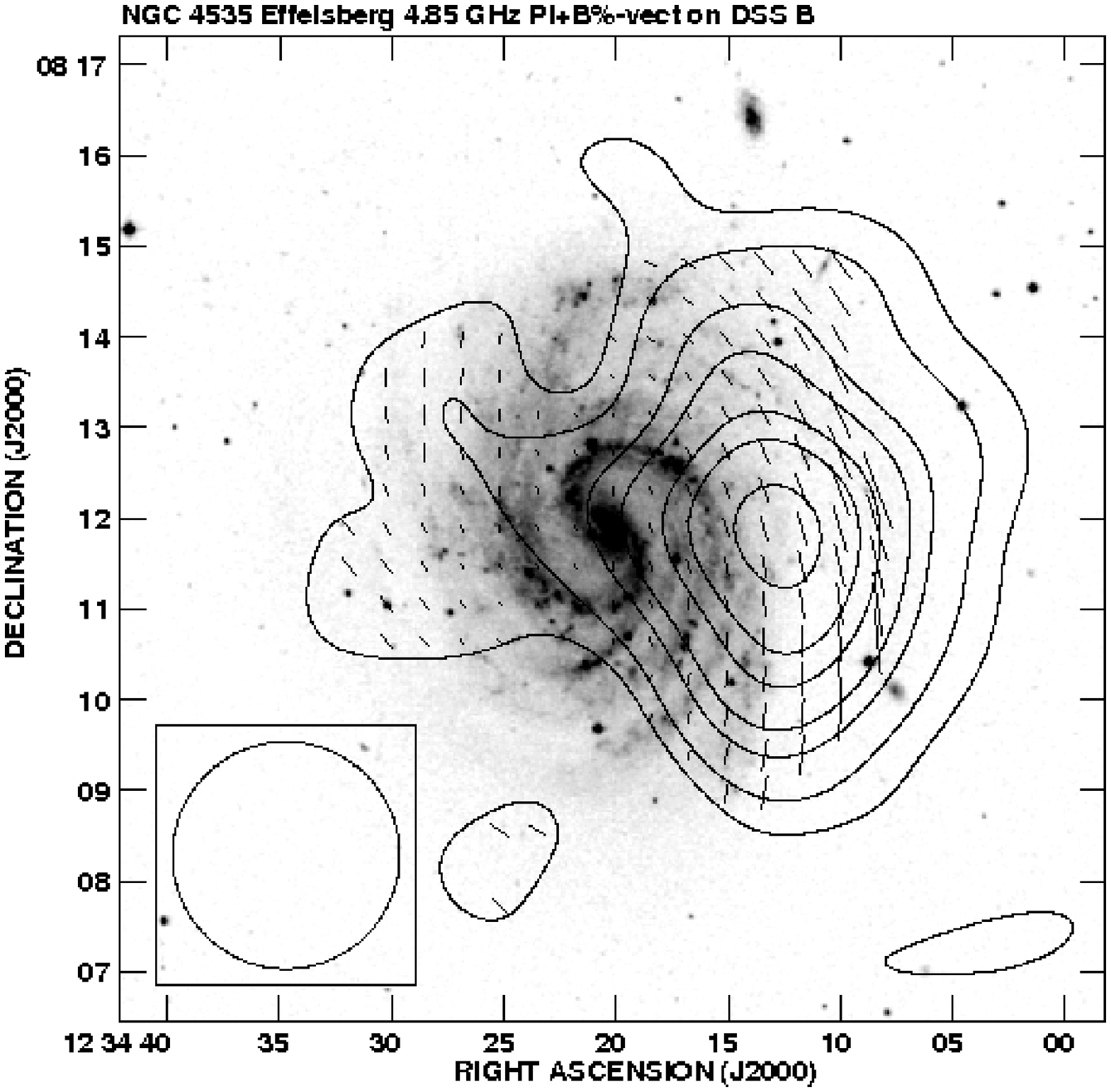}}
 \end{center}
 \caption{ 
The map of polarized intensity of NGC\,4535 at 4.85~GHz with apparent B-vectors 
of polarization degree overlaid onto the 
DSS blue image. The contours are 3, 5, 8, 10, 13, 15, 18 $\times$
0.1~mJy/b.a. and a vector of $1\arcmin$ length corresponds to the
polarization degree of 60\%. The map resolution is $2\farcm 5$. The beam size
is shown in the bottom left corner of the figure.
}
\label{4535pi}
  \end{figure}

The map of polarized intensity reveals a very strong asymmetry with  75\% of the 
polarized flux coming from the western half of the disk (Fig.~\ref{4535pi}). The 
peak of the polarized emission is located outside the optical structure. Polarization 
B-vectors in this structure are generally parallel to the western optical arm of the 
galaxy. As the total power is distributed symmetrically, the asymmetry in the polarized 
intensity is due to a gradient in the polarization degree (See 
Sect.~\ref{compress}). The extension to the northeast is due to a background source.

\subsection{NGC\,4548}

\begin{figure}[ht]
 \begin{center}
  \resizebox{8cm}{!}{\includegraphics{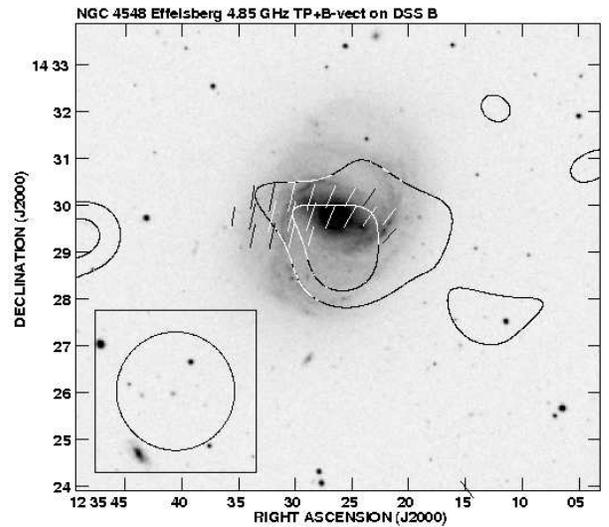}}
 \end{center}
 \caption{ 
The total power map of NGC\,4548 at 4.85 GHz with apparent
B-vectors of polarized intensity overlaid onto the DSS blue
image. The contours are 3, 4 $\times$ 0.77~mJy/b.a. and a
vector of $1\arcmin$ length corresponds to the polarized
intensity of  0.84~mJy/b.a. The map resolution is $2\farcm 5$.
The beam size is shown in the bottom left corner of the figure. 
}
\label{4548tp}
\end{figure}

\begin{figure}[ht]
\begin{center}
\resizebox{8cm}{!}{\includegraphics{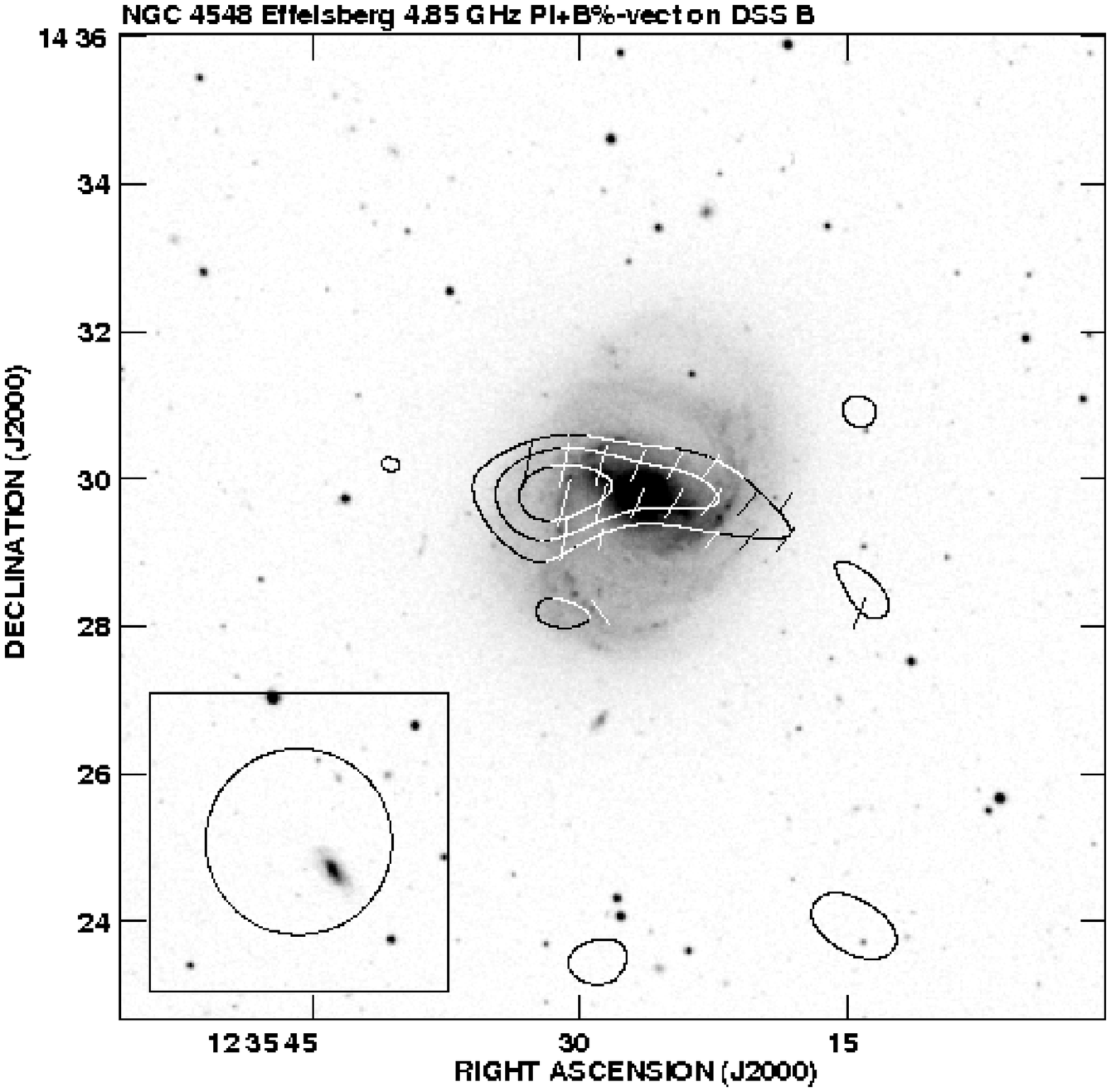}}
\end{center}
\caption{ 
The map of polarized intensity of NGC\,4548 at 4.85~GHz with apparent 
B-vectors of polarization degree overlaid onto the 
DSS blue image. The contours are 3.5, 4.5, 5.5 $\times$
0.09~mJy/b.a. and a vector of $1\arcmin$ length corresponds 
to the polarization degree of 30\%. The map resolution is $2\farcm 5$. The beam 
size is shown in the bottom left corner of the figure.
}
\label{4548pi}
  \end{figure}

NGC\,4548 is located in the northern part of the Virgo Cluster at
the distance of 2.4$\degr$ (0.72\,Mpc in the sky plane) from Virgo\,A. It is an anaemic
galaxy showing little signs of star formation and it is very
poor in neutral gas (Cayatte et al.~\cite{cayatte}). 
It shows a  generally weak total power  emission peaking
south of the optical nucleus (Fig.~\ref{4548tp}). The polarized
intensity is only slightly above the noise level
(Fig.~\ref{4548pi}), but the degree of polarization is quite
high reaching $\approx$ 15\%. The polarized emission
is concentrated in the central parts of the galaxy and extends
along the bar. It coincides with an east-west elongated
hole in the \ion{H}{i} emission (Vollmer et al.~\cite{vollmer3}). 
The polarized emission peaks are on the eastern end of the bar. 
The observed B-vectors are roughly perpendicular to the bar.

\subsection{NGC\,4654}

NGC\,4654 is located in the eastern part of the cluster, 
in the  area of the eastern subcluster (Gavazzi~\cite{gav77}). Its
angular distance  from Virgo~A is $3\fdg 3$ (0.99\,Mpc in the sky plane). The neutral gas distribution
has a comet-like shape (Phookun \& Mundy~\cite{phookun}) with a gaseous tail extending  towards
the southeast. As it is poorly resolved at 4.85~GHz and weak at
10.45~GHz we discuss the radio and polarized structures using
our observations at 8.35~GHz.
 
\begin{figure}[ht]
 \begin{center}
  \resizebox{8cm}{!}{\includegraphics{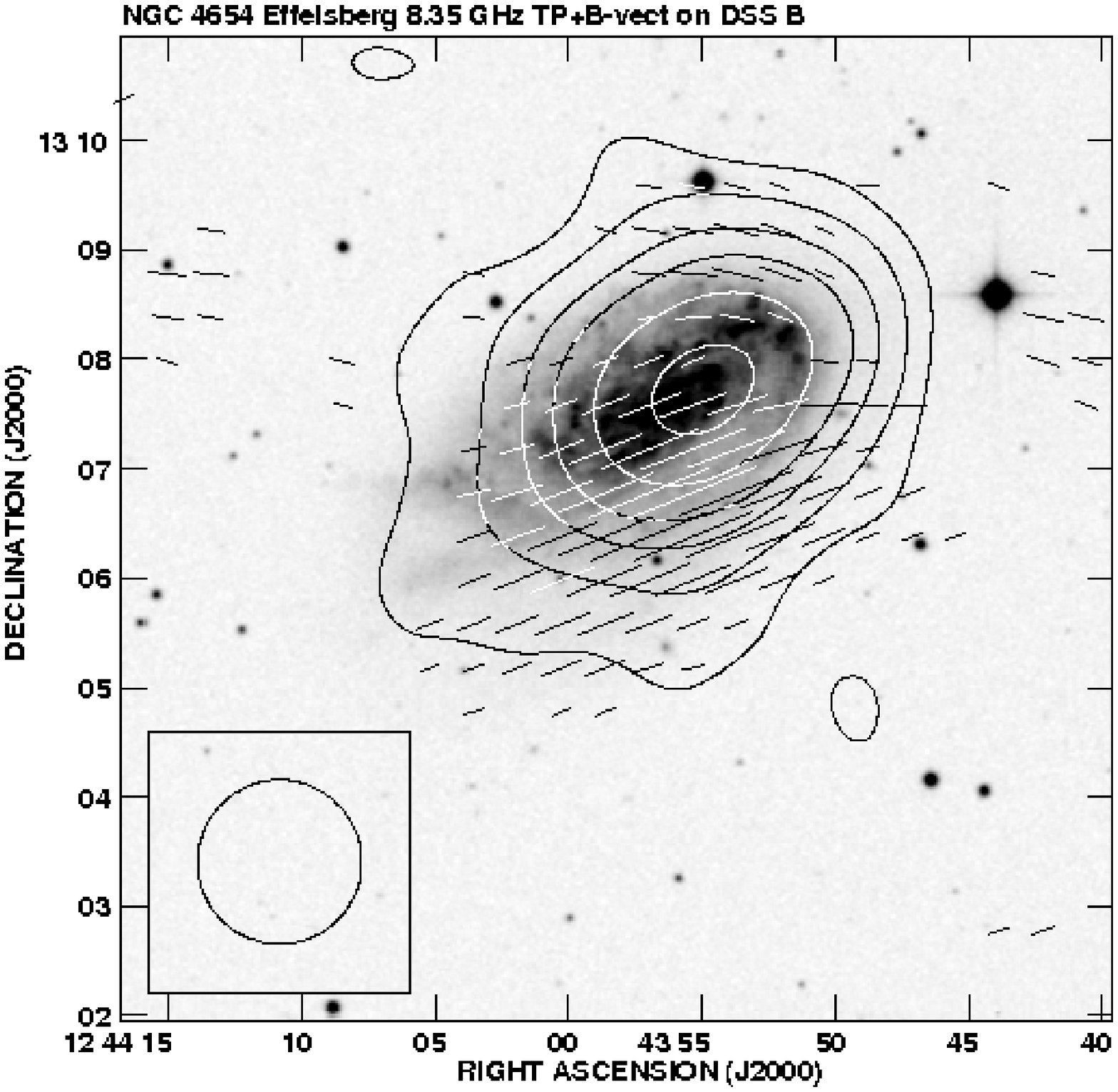}}
 \end{center}
 \caption{ 
The total power map of NGC\,4654 at 8.35 GHz with
apparent B-vectors of polarized intensity overlaid onto the DSS
blue image. The contours are 3, 8, 16, 25, 40, 60 $\times$
0.25~mJy/b.a. and a vector of $1\arcmin$ length corresponds 
to the polarized intensity of 0.4~mJy/b.a. The map resolution is $1\farcm
5$. The beam size is shown in the bottom left corner of the figure.
}
\label{4654tp}
\end{figure}

\begin{figure}[ht]
\begin{center}
\resizebox{8cm}{!}{\includegraphics{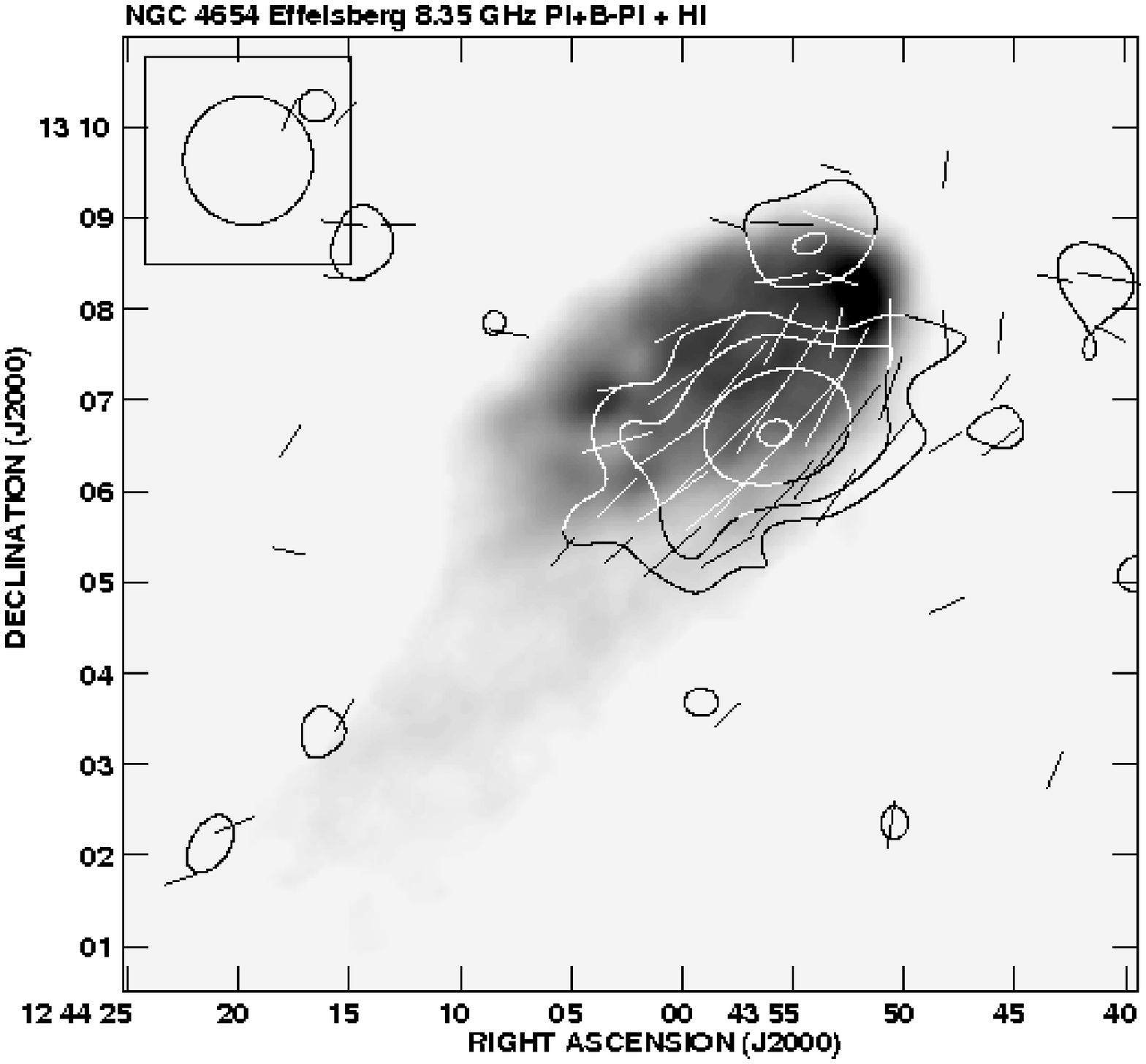}}
 \end{center}
 \caption{ 
The map of polarized intensity of NGC\,4654 at 8.35~GHz with apparent B-vectors 
of polarization degree overlaid onto the \ion{H}{i}
image (Phookun \& Mundy~\cite{phookun}). The contours are 3.5, 5, 8, 13, 18, 21 $\times$
 0.03~mJy/b.a. and a vector of $1\arcmin$ length corresponds 
to the polarization degree of 12.5\%. The map resolution is $1\farcm
 5$. The beam size is shown in the top left corner of the figure.
}
\label{4654pi}
\end{figure}

The distribution of the total power emission from NGC\,4654 
resembles the overall shape of the optical image and seems to 
reflect the star-forming activity in general (Fig.~\ref{4654tp}). 
The peak of total power is slightly shifted towards the optically 
bright clump in the NW side of the galaxy visible in the optical 
images as well as in the \ion{H}{i} maps (see Fig.~\ref{4654pi}). 
The polarized emission has again a highly asymmetric 
distribution. The peak of the polarized brightness is considerably 
shifted from the disk towards the SW (Fig.~\ref{4654pi}). It is  
located at the base of the aforementioned \ion{H}{i} plume extending 
in the SE direction. The B-vectors are parallel to the 
plume instead of following the optical arm which turns up 
northwards in the SE disk.

Recently Soida et al.~(\cite{soidavoll}) presented a high resolution VLA 
map of polarized intensity of NGC\,4654 at  4.85\,GHz.
It confirms the existence of the polarized feature associated 
with the \ion{H}{i} tail. It also
shows a second (weaker) polarization peak in the NW disk region. In our
Effelsberg map having a lower resolution, the latter is
significantly weakened due to beam depolarization effects, as the
magnetic field rapidly changes its orientation in this disk
part.

\begin{table*}[t]
\caption{Integrated data of studied galaxies}
\begin{flushleft}
\begin{tabular}{p{6mm}p{20mm}p{20mm}p{20mm}p{20mm}p{20mm}p{20mm}p{13mm}}
\hline
NGC & {\bf $S_{4.85\rm GHz}$} & {\bf $S_{4.85\rm GHz}$} & {\bf $S_{8.35\rm GHz}$} & {\bf $S_{8.35\rm GHz}$} & {\bf $S_{10.45\rm GHz}$} & {\bf $S_{10.45\rm GHz}$} & \% p\\
    & (TP) [mJy] & (POL) [mJy] & (TP) [mJy] & (POL) [mJy] & (TP) [mJy] & (POL) [mJy] & -- \\
\hline
4438 & 86.8$\pm$6.5 & $5\pm$0.6 & -- & -- & -- & -- & 5.8$\pm$0.8\\ 
\hline
4501& 102.4$\pm$8.1 & 10.5$\pm$1 & -- & -- & -- & -- & 10.3$\pm$1.3$^1$\\
  & -- & -- & -- & -- & 53.5$\pm$5.1 & 6.8$\pm$1.6 & 12.8$\pm$3.3$^3$\\
\hline
4535& 39.5$\pm$4.3 & 4.6$\pm$0.8 & -- & -- & -- & -- & 11.7$\pm$2.3\\
\hline
4548& 5.7$\pm$1.8 & 0.6$\pm$0.3 & -- & -- & -- & -- & 11.2$\pm$6.2\\
\hline
4654& 50.2$\pm$4.4 & 2.2$\pm$0.5 & -- & -- & -- & -- & 4.4$\pm$1$^1$\\
  & -- & -- & 34.9$\pm$3 & 1.4$\pm$0.7 & -- & -- & 4$\pm$2.5$^2$\\
     & -- & -- & -- & -- & 24.3$\pm$3.5 & 1.6$\pm$0.9 & 6.4$\pm$3.8$^3$\\
\hline
\end{tabular}
\label{obdata}
\end{flushleft}
\% p -- polarization degree\\ 
TP = total power flux density, POL = polarized flux density\\
$^1$ at 4.85 GHz\\
$^2$ at 8.35 GHz\\
$^3$ at 10.45 GHz\\
\end{table*}

\subsection{Faraday rotation}
\label{frot}

For both NGC\,4501 and
NGC\,4654 polarization observations at two frequencies were
performed, which enables the  determination of mean Faraday
rotation between 4.85~GHz and 10.45~GHz. The
high-frequency maps were convolved to the resolution at 4.85~GHz.
The intrinsic Faraday rotation in spiral galaxies varies between
positive and negative values which tend to cancel when observed
with a large beam. This allows to detect possible
foreground rotation in the cluster medium. 

In case of NGC\,4501 the Faraday rotation measure (RM) changes 
gradually from +14 to +46~rad/m$^{2}$ across the disk in the 
SE-NW direction, with a mean value of +30~rad/m$^{2}$. In 
NGC\,4654 we see two distinct areas of moderately high rotation
of 57~rad/m$^{2}$ in the SE part of the disk changing to 
-57~rad/m$^{2}$ -- -66~rad/m$^{2}$  on the opposite disk side. The 
mean RM amounts to only -4~rad/m$^{2}$. These extreme
values correspond to a maximum rotation 
of polarization plane by 13\degr\, at 4.85~GHz (either counterclockwise or
clockwise). This amount is in fact smaller for most of our data. 
Several times smaller rotation is expected at higher frequencies. 
For observations with our large beam we can use the apparent 
B-vectors as a good approximation of the sky-projected orientation of 
large scale regular magnetic fields, accurate to some $\pm 10$\degr\, at the lowest of 
our frequencies.

\section{Discussion}

\begin{figure}[ht]
\begin{center}
\resizebox{8cm}{!}{\includegraphics[angle=0]{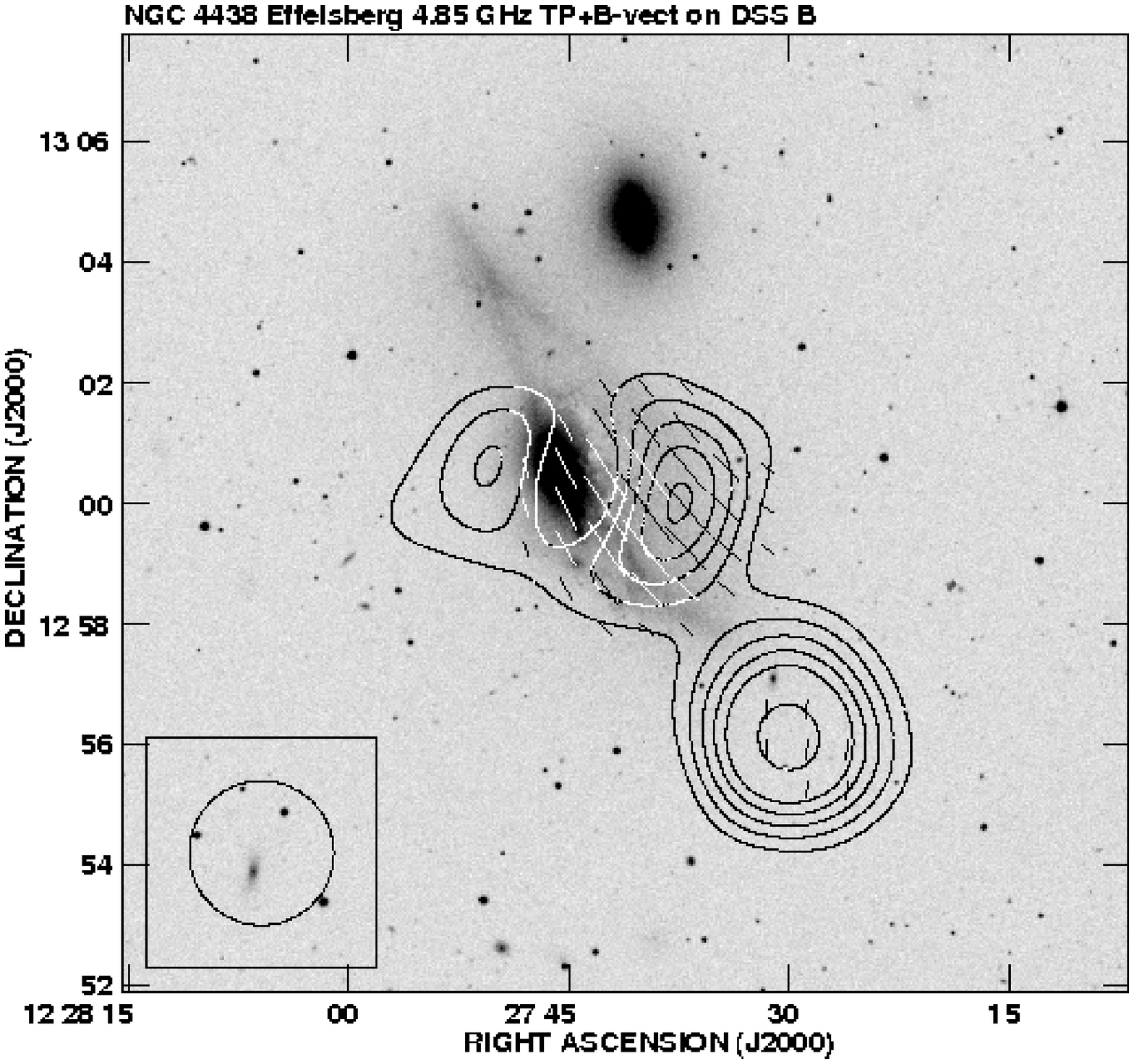}}
\end{center}
\caption{ The residual total power brightness after the "maximum subtraction" of  
the central source with B-vectors for the brightest polarized peak, both overlaid 
onto the blue DSS image. The contour levels are 6, 9, 12, 15, 18, 25 $\times$ 
0.7~mJy/b.a., and the polarization vectors are truncated at the level of 0.5~mJy/b.a.
A vector of $1\arcmin$  length corresponds to 2~mJy/b.a.
}
\label{4438res}
\end{figure}

The results shown in Section~3 present a variety of effects
occurring in cluster galaxies. Below, we make an attempt to
categorize various effects according to different physical
processes governing the interactions.

\subsection{Galactic anaemia}

NGC\,4548 (Figs.~\ref{4548tp} and \ref{4548pi}), 
the only anaemic galaxy in our sample has no close
companions and shows an \ion{H}{i} deficiency of 0.76 (Koopmann
\& Kenney~\cite{koopman}, Cayatte et al.~\cite{cayatte2}). 
It is a very weak radio source. The integrated flux density
(hence total luminosity) at 4.85~GHz is by a factor of 6 -- 10
times smaller than that of other spirals in our sample. 
It is also several times smaller than the total flux density of
54~mJy (roughly half of it coming from the disk region - Gallimore et  al.~\cite{gallimore})
measured at this frequency by Becker et al.~\cite{becker} 
for NGC\,4579 -- another Virgo Cluster object of similar
morphological type and \ion{H}{i} deficiency (Cayatte et 
al.~\cite{cayatte2}). This makes NGC\,4548 an exceptionally
weak radio emitting galaxy, even for an anaemic spiral.

\subsection{Ram pressure effects - compression}
\label{compress}

Otmianowska-Mazur \& Vollmer~(\cite{n4254}) show that the
magnetic perturbations in the outer disk may last 0.7 -- 0.9
Gyr after the passage through the dense ICM. Within this time
galaxies travelling at the speed of over 1000~km/s with
respect to the ICM (many Virgo Cluster spirals move faster) might
have moved by almost 1 Mpc (considerably more than 3\degr\, in
case of transverse motions) from the cluster core. Thus, the
observed signatures of magnetic field compression may either have
occured recently or they may constitute a sensitive, long 
lasting memory of the epoch (some hundreds Myr ago) when the
galaxy was much closer to the Virgo Cluster centre. These two
scenarios are presently difficult to discriminate.

There are clear examples of galaxies in our sample that show
primary importance of ram pressure effects, without any
influence of tidal interactions. The first of them is  
NGC\,4501 (Figs.~\ref{4501tp} and ~\ref{4501pi}), in which the
asymmetry of polarized emission is significant with more than
85\% of the polarized emission found southwest of the major axis.
This is unusual for normal spirals and may result from the
enhancement of regular and/or random but anisotropic magnetic
fields by compression as described in detail by Beck et
al.~(\cite{beck05}). If the enhancement results from compression
in the sky plane, both these magnetic field structures may yield
high polarization. We note that the discrimination between
unidirectional and anisotropic random magnetic fields needs
Faraday rotation data with an appropriate resolution.

Our beam is too large to decide whether we are dealing
with a narrow compressional ridge here. We note however that the
maximum of the polarized flux density occurs on the disk side
where strong gradients of \ion{H}{i} distribution were observed
(Cayatte et  al.~\cite{cayatte}) and a bright spot of X-ray
emission can be seen (B\"ohringer et al.~\cite{bohringer}).
Moreover, the optical structure also shows a smaller pitch angle
on this disk side than on the opposite side. This asymmetry
resembles that in NGC\,4254, in which compression effects were
suggested (Chy\.zy et al.~\cite{chyzy}). The latter object, in 
which the VLA data reveal a narrow
compressional ridge, shows only a global asymmetry of the
polarized  brightness when observed at resolution similar to that
for NGC\,4501 (Soida et al.~\cite{soida}). The explanation of the
asymmetry of NGC\,4501 by ram pressure stripping is supported by
its underabundance of \ion{H}{i}. The deficiency parameter of 0.34  
(Koopmann \& Kenney~\cite{koopman}) places this galaxy among objects 
suffering significant stripping effects.

The \ion{H}{i} data by Cayatte et al.~(\cite{cayatte}) indicate
that the NW half of the disk of  NGC\,4501 is approaching us. If
the spiral is trailing this would mean that the  compressed edge
is the remote one and the galaxy is viewed "from below" (see
also Onodera et al.~\cite{onodera}). The compression along the
remote edge agrees well with a high positive velocity of
NGC\,4501 with respect to M\,87 and to the average velocity for
other galaxies in this Virgo Cluster region (LEDA database). The
excess of polarization occurs in the "head-on" disk side and can
be attributed to a compressional enhancement of regular or random 
and anisotropic magnetic fields.

Another example of ram pressure effects is NGC\,4535
(Figs.~\ref{4535tp} and \ref{4535pi}), which has symmetric, 
grand-design distribution of stars. In this case the peak of
polarized intensity is located {\em outside} the bright optical
spiral arms, while the total power emission is as symmetric as
the optical structure.   
The asymmetry in the distribution of
polarized intensity is clearly caused by an E-W gradient in the
degree of magnetic field ordering. In fact, the mean
polarization degree (integrated in areas of roughly
$5\arcmin\times 5\arcmin$ located east and west of the N-S line 
going through the centre) is only 6\% in the eastern disk half
rising to 16\% in the western one. This asymmetry is surprising 
because NGC\,4535 is located in the southern Virgo extension
(B\"ohringer et al.~\cite{bohringer}), where the intracluster gas
density is rather moderate. As noted in the beginning of
this section, we cannot exclude the possibility that the
distortion may have occurred some hundreds of Myrs ago (see
Otmianowska-Mazur \& Vollmer~\cite{n4254}). In this case strong
magnetic anomalies accompanying only weak perturbations observed
in the \ion{H}{i} gas (Cayatte et al.~\cite{cayatte}) may imply
that the magnetic fields provide a very long-lasting memory of
past interactions,  possibly caused by a back-falling gas
stripped some hundreds of Myrs ago. On the other hand, if
NGC\,4535 is being stripped now, it may be moving in the sky
plane from east towards the west. This information is
impossible to acquire from radial velocity studies alone. The
latter possibility would also mean that the magnetic field
structure may respond strongly to even quite weak ram pressure
efects.

\subsection{Tidal interactions mixed with ram pressure effects}

Gravitational encounters may play a significant role in the
evolution of cluster spirals as the spatial density of objects
is high. However, in the cluster centre disruptive encounters
are rare as high relative velocities of galaxies make the
encounters very short; as shown by numerical simulations (Soida
et al.~\cite{soidavoll}, Vollmer et al.~\cite{ vollmer4}) typical
timescales for developing the observed signatures of tidal
interactions (with some admixture of ram pressure) involve about
500 Myr for loose encounters in the Virgo Cluster outskirts and
about 150 Myr close to the Cluster core. Strong gravitational
effects may result in enhanced nuclear activity (either LINER,
Seyfert-type or nuclear starburst) triggered by highly
non-circular motions. General distortions of the structure may
occur, and those involving the stellar component may constitute
an unambiguous signature of tidal effects. At the same time
perturbations of the gaseous disk and of the magnetic field
structure may develop, being often accompanied by
magnetized gaseous tails which might provide additional
effects. There are two galaxies in our sample that show
evidence for tidal perturbations as well as, they reveal
ram pressure effects exerted by the cluster medium. We discuss
these cases in detail.

\subsubsection{NGC 4438}
\label{4438d}

The asymmetrically disturbed optical morphology of NGC\,4438 as well as
observations in CO (Combes et al.~\cite{combes}, Vollmer et 
al.~\cite{vollmer4}), H$\alpha$ (Kenney et al.~\cite{kenney}, 
Chemin et al.~\cite{chemin4438}) and X-rays
(Machacek et al.~\cite{machacek}) clearly suggest tidal
interactions with the companion galaxy NGC\,4435. Furthermore,
the X-ray properties of the gas outside the nuclear regions of these galaxies
trace their past collision (Machacek et al.~\cite{machacek}). 

NGC\,4438 is the only galaxy in our small sample showing a strong 
central source; as it is also of earliest Hubble type (Sa), it
has the strongest mass concentration (hence deepest gravitational 
potential minimum) in the centre. The VLA A and B-array
observations of Hummel \& Saikia (\cite{hummel}, resolution
0\farcs4 -- 2\arcsec) show that about 30\% of the total flux
density at 4.86~GHz comes from an arcsecond-scale compact radio
shell surrounding the nucleus, classified as a LINER (NED
database). They also detected radio extensions (up to 10\arcsec)
perpendicular to the galaxy plane. Such a nuclear activity can
possibly be triggered by tidal encounters. Our observations
indeed reveal a strong total intensity peak at the position of
the galaxy centre (Fig.~\ref{4438tp}). The recent D-array
VLA map at 4.86~GHz  by Vollmer et al.~(\cite{letter}) has a lower
resolution of  18\arcsec \, but recovers almost all the extended
flux seen by us with the Effelsberg radio telescope. This map
shows that all structures inside central 18\arcsec \, of
NGC\,4438 account for no more than 50\% of the total galaxy flux
density detected in our single dish data.

The map size (20\arcsec) in Hummel \& Saikia is too small to
observe the (probably extended) polarized peak seen in our single
dish observations, shifted by almost $1\arcmin$ westwards from
the centre. Our resolution is too low to definitely decide whether
the shift has an external origin - either tidal interactions or
ram pressure effects. The "maximum subtraction method"
(Chy\.zy et al.~\cite{chyzy6822}) reveals a strong total power
residual (peak of $\approx$ 15~mJy/b.a.) SW of the nucleus,
displaced from the centre in the same direction (but 
slightly farther out) as the polarized
blob (Fig.~\ref{4438res}). We checked that it is not due to a
(possibly) non-Gaussian Effelsberg beam, and is not a sidelobe
effect. A residual total power peak has been found in the same
place by Kotanyi
\& Ekers~(\cite{kotanyi}) who claimed it to be either due to
external compression or tidal interactions. Our residual
source is thus real, polarized with B-vectors parallel to the
disk, and may be associated with extraplanar dust, H$\alpha$
and X-ray structures.

The total power residual possibly associated with the
polarized peak that has a {\em disk-parallel} magnetic field
might be explained by cluster ram pressure compressing the
magnetic field. On the other hand, the tidal processes 
could stretch the magnetic field and enhance the polarized emission
as well as align magnetic fields with extraplanar optical
filaments parallel to the disk. Sensitive polarimetric
observations with a resolution of $\approx$ 15\arcsec\, --
20\arcsec\, are necessary to evaluate contributions from both
processes.

With the high sensitivity to extended structures our polarization
map also reveals weak extended emission on both sides of the disk
extending  up to 5\arcmin\, and proven to be not due to 
instrumental polarization (Sect.~\ref{4438}). These polarized
features have an integrated intensity of 1.1$\pm$0.1 mJy and are
associated neither with optical filaments nor known background
sources. The magnetic field vectors have different orientation
here than in the disk region or in the strong western extension,
and are highly inclined to the disk. 

If these extensions are the large-scale magnetized outflows
they must have a pressure and energy density high enough to
overcome the ram pressure in the surrounding cluster medium.  As 
suggested by Vollmer et al.~(\cite{vollmer4}) the ambient kinetic
pressure in the vicinity of NGC\,4438 is $\simeq 10^{- 12}$g\,
cm$^{-1}$s$^{-2}$. Though our resolution is modest  an assumption
that the extraplanar radio features fill our beam can provide the
minimum possible values of magnetic and cosmic ray pressures.
Assuming the condition of pressure balance between the magnetic
field and cosmic rays we obtain the minimum strength of a stable
magnetic field in the outflows to be 3 -- 3.5$\mu$G. Thus the
joint pressure of magnetic fields and cosmic rays in these
features must be at least of order of  $ 10^{-12}$g\,
cm$^{-1}$s$^{-2}$ (likely to be considerably higher). For these
calculations we adopted the low energy cosmic ray cutoff at 300~MeV, 
the nonthermal spectral index of 1.0 (similar to that in
outflows in NGC\,4569, Chy\.zy et al.~\cite{chyzy2}) and a
pathlength of some 10~kpc along the line of sight. If the
extraplanar structures are better collimated the magnetic and
cosmic ray pressures become even several times higher. Thus the
outflows have enough power to overcome the intracluster gas 
pressure. Sensitive Effelsberg observations at higher frequencies
are desirable to determine the detailed structure and the
spectral index of these possible outflows and thus, to 
discover their physical nature.

\subsubsection{NGC 4654}
\label{4654}

Another galaxy in which tidal forces may at least play some role
is NGC\,4654, which is probably interacting with the neighbouring
NGC\,4639. However, the cometary shape of its \ion{H}{i}
distribution and the gaseous tail extending towards the SW  
may be caused by ram pressure effects as the galaxy moves through 
the ICM (Phookun \& Mundy \cite{phookun}). The galaxy shows no
strong central source, probably because of less pronounced
central mass concentration as expected for its much later type
(Sc) than NGC\,4438. We note that the B-vectors clearly follow
the tail (as expected for its outflowing motion), rather than the
spiral pattern in this part of the disk. Recent simulations of
the magnetic field of this galaxy (Vollmer et 
al.~\cite{vollmer5}, confirmed by Soida et al.~\cite{soidavoll})
suggest that a mixture of tidal interaction with some ram
pressure effects is the best explanation of the radio
polarization structure of NGC\,4654. In particular, the bright
polarized blob in the western part of the gaseous tail with the
magnetic field stretched along the tail as well as, the weaker
polarization spot NW of the centre are well reproduced.

\subsection{Global properties}
\label{global}

\begin{figure}[ht]
 \begin{center}
  \resizebox{8cm}{!}{\includegraphics[angle=-90]{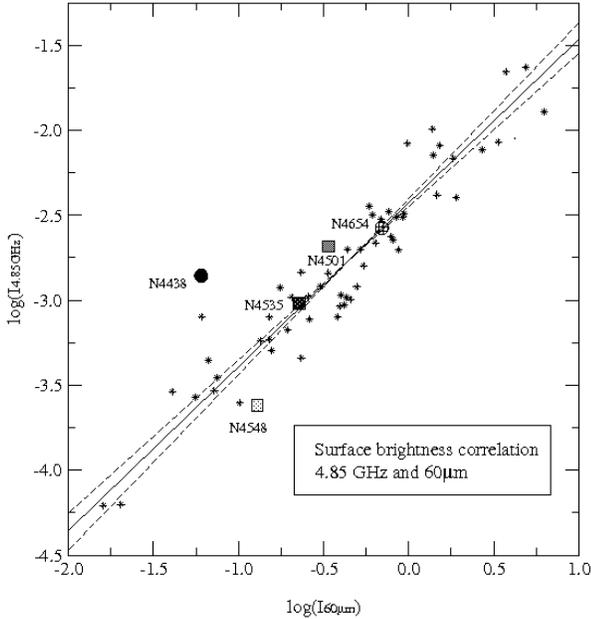}}
 \end{center}
 \caption{
The radio - FIR correlation diagram for our Virgo objects plotted as symbols with
labels and for the reference sample of  galaxies observed by Gioia et
al.~(\cite{gioia}) with an extension towards low-surface 
brightness objects observed by Chy\.zy et al.~(\cite{chyzy06}) --
both as dots. The surface brightness  at 4.85~GHz (Jy/$\sq$\arcmin) and at
$60\mu$m ((Jy/$\sq$\arcmin, see text) is used. The solid  curve is an orthogonal fit to
reference galaxies with a slope of 0.96$\pm$0.06.  The dashed
lines show the ``regression scissors'': maximum and minimum slope 
(1.03 and 0.90) allowed by the data scatter. 
}
\label{firc}
\end{figure}

To compare the integrated total power emission from the Virgo 
Cluster spirals to non-cluster nearby spiral galaxies we analyze
the radio -- far-infrared (FIR) correlation. As statistical
variables we used the distance-independent  quantities: the radio
and FIR surface brightness. They are computed by dividing the
integrated flux density at 4.85~GHz and at $60\mu$m respectively
by their face-on corrected observed disk surface. For that
purpose we used the extinction-corrected diameters from the LEDA
database. As a homogenous reference sample we took the galaxies
measured at 4.8~GHz by Gioia et al.~(\cite{gioia}) who used the
same instrument and a similar method of integrating the total
power flux density. The reference sample has been expanded
towards low intensities using the 4.85~GHz Effelsberg
measurements of low surface brightness galaxies (Chy\.zy et
al.~\cite{chyzy06}). The flux densities at $60\mu$ were taken
from Helou \& Walker (\cite{iras1}) and if not present there,
from Moshir et  al.~(\cite{iras2}).

The radio-FIR correlation for our Virgo sample and the reference
sample is shown in  Fig.~\ref{firc}. Three normally star-forming
galaxies: NGC\,4501,  NGC\,4535 and NGC\,4654, closely agree with
the relation derived for non-Virgo spirals, which has a slope of 
0.96$\pm$0.06. In contrast, NGC\,4438 lies considerably above 
this relation. To check for the significance of this deviation we 
computed the r.m.s. deviation of the radio brightness from the 
orthogonal fit shown in Fig.~\ref{firc}. We obtain 0.17 
log~(Jy/$\sq$\arcmin), thus NGC\,4438 deviates significantly by
4.4$\sigma$ from the correlation for normal spirals. The
deflection corresponds to an excess of radio over FIR emission
by a factor of about 5 -- 5.5. An abnormally high radio/FIR
ratio has also been found for this galaxy by Niklas et
al.~(\cite{nikvir}). A similar enhancement of radio emission in
Coma Cluster spirals was attributed to compression of magnetic
fields caused by the ambient medium by Gavazzi~(\cite{gavaz3}). 

The excess could be due to either the western polarized radio
peak (Fig.~\ref{4438res}), probably associated  with some
extraplanar radio emission (Kotanyi \& Ekers~\cite{kotanyi}), or
to nonthermal structures unrelated to  star formation originating
in the nuclear region. According to the VLA map at 4.85~GHz by
Vollmer et  al.~(\cite{letter}) the central region (innermost 
18\arcsec) of NGC\,4438 contributes less than 50\% to the total
flux density at 4.85~GHz while the images from the ISO archive
show this region to be the source of the bulk of FIR emission. On
the other hand the extraplanar radio source of Kotanyi \& Ekers
(\cite{kotanyi}) emitting much less in FIR amounts to only some
10 -- 12\% of the total flux at 1.4~GHz, probably the same
fraction at 4.85~GHz. Altogether this is too little to be an excess of a
factor of 5, thus there could be another component undetected by
interferometers and not related to star formation, contributing
to the excess of radio over FIR emission. To establish a possible
association of the excess with circumnuclear or extraplanar
structures or to say whether we are dealing with compressional
enhancement of magnetic fields as proposed by Reddy \& Yun
(\cite{reddy}) we need a substantially better resolution together with
enough sensitivity to extended structures. We note finally that a
similar excess of radio over FIR emission for cluster spirals
has been noted by Gavazzi~(\cite{gavaz3}).

NGC\,4548 has a radio surface brightness lower than any galaxy 
in the sample of Gioia et al.~(\cite{gioia}). It also belongs to 
the weakest FIR emitters in that sample. Compared to another
radio-bright Sb type Virgo Cluster spiral, NGC\,4501, the
discussed galaxy is redder both in $(U-B)_c$  and $(B- V)_c$
(extinction corrected values from LEDA) by about 1\fm 2. It is 
located at the extreme red end of the U-B -- B-V colour
distribution of more than 120 Sb field galaxies brighter than
14\fm 5 selected from the LEDA database. NGC\,4548 is obviously
deficient in massive stars producing cosmic rays electrons
(responsible for the radio emission) via supernova explosions.
Its weak radio emission compared to normal spirals is therefore
largely due to a low star formation level caused by the 
"anaemia", causing its low position on the radio-FIR correlation.

Additionally, NGC\,4548 shows some weak {\em negative}
deviation from the correlation line for normal spirals
(Fig.~\ref{firc}). Thus, there is some evidence that the radio
emission is for its FIR brightness even weaker than expected.
However, the deviation amounts to -1.9$\sigma$ of the regression
residuals. Therefore, this anaemic galaxy shows only weak evidence
for an additional deficit of radio emission (hence of total
magnetic fields) regarding to what is expected from its low
star formation level.

\subsection{Intracluster Faraday rotation}
\label{rm}

The measurements of Faraday rotation of a diffuse emission
from the Perseus Cluster (de Bruyn \& Brentjens~\cite{bruyn})
or of background sources shining through a number of Abell
clusters (Clarke et al.~\cite{clarke}) indicate the existence of
microgauss regular magnetic fields in the intracluster medium.
They yield Faraday rotations reaching $\pm$40 rad/m$^2$ out to
projected distances of 0.5 -- 1 Mpc from the cluster core. Thus,
the Faraday rotation effects observed by us may be at least
partly due to the intracluster medium. Too far-reaching
conclusions from two objects only are rather dangerous nevertheless,
we note that the distance modulus of NGC\,4501 determined by
Gavazzi et al.~(\cite{gav77}) places it at the mean distance of the
surrounding part of the Virgo Cluster while NGC\,4654 lies
in the front of the eastern subcluster. This would mean
that NGC\,4501 may shine through at least half of the thickness
of the magnetized intracluster gas while NGC\,4654 lies in the
foreground of the cluster medium. However it is too early to
conclude that the differences in mean Faraday rotation (Sect.~\ref{frot}) reflect the
position of the two galaxies with respect to the magnetized intracluster
gas. The fact that the above estimate of the distance to
NGC\,4654 became questioned by Soida et al.~(\cite{soidavoll})
raises even more doubts. Such questions will be a subject of future
observational work using a much larger sample also including
background objects.

\section{Summary and conclusions}

We present results of the first systematic study of magnetic field structures 
of Virgo Cluster spirals. We used the Effelsberg radio telescope at 
4.85~GHz to detect weak extended total power and polarized 
emission, while the radio-brightest objects were studied in more detail at 
8.35~GHz or at 10.45~GHz. The results are as follows:
 
\begin{itemize}

\item[-] Ram pressure effects influence mainly the gaseous
component. NGC\,4501, located in a relatively dense hot gas
environment, shows a typical wind-swept \ion{H}{i} asymmetry (and
some weak optical effects) while NGC\,4535, surrounded by a more
dilute hot medium, lacks clear signs of interaction in the
optical domain and in the neutral gas distribution.
Nevertheless, both objects show a strong concentration of
polarized emission on one disk side. In NGC\,4501 it agrees with
\ion{H}{i} asymmetry and with the direction of movement of the
galaxy with respect to the cluster. In case of NGC\,4535  a
strong asymmetry of the magnetic field structure occurs at a much 
lower level of ram pressure effects, providing a clue to the 
galaxy's motion in the sky plane. Alternatively, NGC\,4535 may
still ''remember'' past stripping events.   

\item[-] Two galaxies with obvious signs of tidal interactions 
mixed with ram pressure effects behave in a different way: the
Sa-type NGC\,4438 shows a polarized peak shifted away from the
disk towards the west, situated in a region suspected of
external compression and associated with extraplanar emission in
H$\alpha$ and X-rays. The galaxy has a strong, compact central 
source with clear signs of low-surface brightness magnetized
outflows possibly connected to it (detectable only with a
single-dish instrument), reaching some 25\,kpc away from the
galaxy plane. Another object less affected by ram pressure 
effects, the Sc-type spiral NGC\,4654 has no central source but 
shows a magnetized \ion{H}{i} tail which is best modeled by a 
combination of tidal forces and ram pressure effects. 

\item[-] The lack of gas and a low star formation level in the
anaemic galaxy NGC\,4548 results in weak radio emission, though
this may be not the case for other similar anaemic spirals of a
similar Hubble type (e.g. NGC\,4579). 

\item[-] Three out of five galaxies follow well the radio
(4.85~GHz) -- far infrared  (60$\mu$m) correlation derived for
non-cluster spirals. In contrast, NGC\,4438 shows a strong excess
of nonthermal radio emission. The circumnuclear structures are 
insufficient to explain this phenomenon, but the compressional
enhancement of magnetic fields is a likely cause. On the other
hand, the anaemic spiral NGC\,4548 shows a marginally significant
total power deficiency compared to its FIR emission. 

\item[-] Faraday rotation measurements are only available for two
galaxies: NGC\,4501 and NGC\,4654. The former is probably
located deeper inside the intracluster gas and has a definitely
higher mean Faraday rotation than the latter known to lie in
the front of its local subcluster. To attribute this effect to
the intracluster magnetic fields, a considerably larger sample 
of cluster galaxies and background objects is needed.

\end{itemize}

Despite a modest resolution the ``magnetic diagnostics'' was proven
to trace the effects of interactions with other galaxies or with the
intracluster gas by providing a very long-lasting memory of past interactions 
even when effects are weakly visible in other
species. Single-dish observations were particularly helpful in case of
radio-weak outflows in NGC\,4438. Having in mind promising
results of this study we plan for the near future more systematic
studies involving larger samples of cluster spirals and background
sources.

\begin{acknowledgements}
Three of us: MU, MS and KCh are indebted to Professor Richard Wielebinski 
from Max-Planck-Institut f\"ur Radioastronomie (MPIfR) in Bonn for his multiple 
invitations to MPIfR, where a large part of this work has been done. 
We thank Dr Elly Berkhuijsen from Max-Planck-Institut f\"ur Radioastronomie (MPIfR) in Bonn
for her valuable comments and critical remarks. We would also like to thank the referee, Michael
Dumke, for his substantial remarks. 
This work was supported by the Polish Ministry of Science and Higher
Education, grant 2693/H03/2006/31, and by the Polish-French (ASTRO-LEA-PF) 
cooperation program. We acknowledge the usage of the HyperLeda database
(http://leda.univ-lyon1.fr). The research has also used the NRAO
VLA Sky Survey (NVSS).
\end{acknowledgements}

\end{document}